\documentclass[aps,pre,twocolumn,superscriptaddress,floatfix]{revtex4}
\usepackage{dcolumn}
\usepackage{amsmath}
\usepackage{amssymb}
\usepackage{amsthm}
\usepackage{graphicx}
\usepackage{times}
\usepackage{bm}
\usepackage[T1]{fontenc}
\usepackage{color}
\usepackage{url}
\usepackage[bf]{subfigure}
\usepackage{rotating}
\usepackage{scalefnt}
\usepackage{multirow}

\usepackage{scalefnt}

\theoremstyle{definition}

\newcommand{\A}{\mathbf{A}}
\newcommand{\V}{\mathbf{v}}
\newcommand{\Y}{\mathbf{y}}

\DeclareMathOperator{\I}{\mathbf{I}}

\newcommand{\vertii}[1]{{\left\vert\kern-0.25ex\left\vert #1 \right\vert\kern-0.25ex\right\vert}}

\begin{document}

\title{Universality of eigenvector delocalization and the nature of the SIS phase transition in multiplex networks}

\author{Guilherme Ferraz de Arruda}
\affiliation{ISI Foundation, Via Chisola 5, 10126 Torino, Italy}

\author{J. A. M\'endez-Berm\'udez}
\affiliation{Instituto de Ci\^{e}ncias Matem\'{a}ticas e de Computa\c{c}\~{a}o, Universidade de S\~{a}o Paulo - Campus de S\~{a}o Carlos, Caixa Postal 668, 13560-970 S\~{a}o Carlos, SP, Brazil}
\affiliation{Instituto de F\'isica, Benem\'erita Universidad Aut\'onoma de Puebla, Apartado Postal J-48, Puebla 72570, Mexico}

\author{Francisco A. Rodrigues}
\affiliation{Departamento de Matem\'{a}tica Aplicada e Estat\'{i}stica, Instituto de Ci\^{e}ncias Matem\'{a}ticas e de Computa\c{c}\~{a}o, Universidade de S\~{a}o Paulo - Campus de S\~{a}o Carlos, Caixa Postal 668, 13560-970 S\~{a}o Carlos, SP, Brazil}

\author{Yamir Moreno}
\affiliation{Institute for Biocomputation and Physics of Complex Systems (BIFI), University of Zaragoza, 50018 Zaragoza, Spain}
\affiliation{Department of Theoretical Physics, University of Zaragoza, 50009 Zaragoza, Spain}
\affiliation{ISI Foundation, Via Chisola 5, 10126 Torino, Italy}

\begin{abstract}
Universal spectral properties of multiplex networks allow us to assess the nature of the transition between disease-free and endemic phases in the SIS epidemic spreading model. In a multiplex network, depending on a coupling parameter, $p$, the inverse participation ratio ($IPR$) of the leading eigenvector of the adjacency matrix can be in two different structural regimes: (i) layer-localized and (ii) delocalized. Here we formalize the structural transition point, $p^*$, between these two regimes, showing that there are universal properties regarding both the layer size $n$ and the layer configurations. Namely, we show that $IPR \sim n^{-\delta}$, with $\delta\approx 1$, and revealed an approximately linear relationship between $p^*$ and the difference between the layers' average degrees. Furthermore, we showed that this multiplex structural transition is intrinsically connected with the nature of the SIS phase transition, allowing us to both understand and quantify the phenomenon. As these results are related to the universal properties of the leading eigenvector, we expect that our findings might be relevant to other dynamical processes in complex networks.
\end{abstract}

\maketitle

Universality is at the core of physics \cite{Stanley1971,Marro1999}. Universal properties do not change from one system to another but represent an entire class of them. They allow us to go beyond the observation of macro variables towards the understanding of the mechanisms that trigger a given behavior. Another notable consequence of universality is that by understanding the behavior of one system, we are able to make conclusions about other systems of the same class or governed by the same set of symmetries. Universality in multilayer networks was firstly explored in~\cite{Bermudez2017}, where the eigenvector properties of the corresponding adjacency matrix were shown to follow a simple scaling law. 
Complementarily, the spectral properties of multiplex networks have been recently explored in~\cite{Cozzo2014, deArruda2017Feb, cozzo2018multiplex}. In addition, Ref~\cite{deArruda2018} reported a non-trivial relationship between the eigenvalues of a relatively simple multiplex network composed by two layers. However, the analysis carried out in \cite{deArruda2018} mainly focused on the eigenvalues and their bounds rather than eigenvectors, which may provide additional valuable information about the network structure.

From the dynamical viewpoint, the concept of layer-localization in multiplex networks was introduced in~\cite{deArruda2017Feb}. That is, when a disease is on top of a multiplex network, it can be localized in one or a subset of layers. This phenomenon depends on the intra-layer configurations and also on the coupling strength between layers. Moreover, it is intrinsically linked to the localization properties of the eigenvectors of the network adjacency matrix, commonly measured by the inverse participation ratio, $\text{IPR}$. Although this phenomenon was well characterized in~\cite{deArruda2017Feb}, the mechanism driving it was not fully understood. In more technical words, the leading eigenvector of a multiplex network can be in one of two different regimes as a function of the coupling parameter between layers: layer-localized regime and delocalized regime. However, the precise definition of the structural transition between those regimes, to the best of our knowledge, is still lacking in the literature. Therefore, here we propose a definition for the transition point between layer-localization to delocalization, showing that it can be used to collapse the $\text{IPR}$ curves in a wide range of network configurations. This collapsing also reveals the universality of the transition. Finally, as an application, we analyze the disease spreading on multiplex networks, providing a dynamical condition for the layer-localization to delocalization phase transition.

In its more general form, multiplex networks are composed by $m$ layers \cite{Kivela2014,BoccalettiPR2014,Bianconi2018, Aleta2019}. Each layer has at most $n$ nodes which might have a counterpart in the other layers. Here, we restrict to 2-layer multiplex networks
(i.e.,~$m=2$) where each layer has $n$ nodes and each node has a counterpart on the other layer. 
Formally, these networks can be represented by the adjacency matrix $\A$ whose eigenvalue problem is 
given as 
\begin{equation} \label{eq:definition}
\left[
\begin{array}{c|c}
\A_1 & p\I \\
\hline
p\I & \A_2
\end{array}
\right]
\left[
\begin{array}{c}
 \V_1 \\
 \hline
 \V_2
\end{array}
\right] =
\lambda
\left[
\begin{array}{c}
 \V_1 \\
 \hline
 \V_2
\end{array}
\right] =
\lambda \V = \A \V,
\end{equation}
where $\A_{1,2}$ are the individual adjacency matrices, $\V_{1,2}$ are the respective sub-vector components, $p$ is the coupling weight, and $\parallel \V \parallel = 1$. 
Furthermore, we focus on the case where there is layer dominance~\cite{Cozzo2013Nov, cozzo2018multiplex}, i.e.,~$\lambda_1 \gg \lambda_2$, where $\lambda_1$ and $\lambda_2$ are the leading eigenvalues of the individual layers.
As a consequence, the components of the leading eigenvector of $\A_{2}$ should be 
relatively small, i.e.,~$(\V_2)_j \approx 0$. This can be easily seen using perturbation theory, see 
e.g.~\cite{Cozzo2013Nov}. 

\begin{figure}[t!]\centering
\includegraphics[width=0.9\columnwidth]{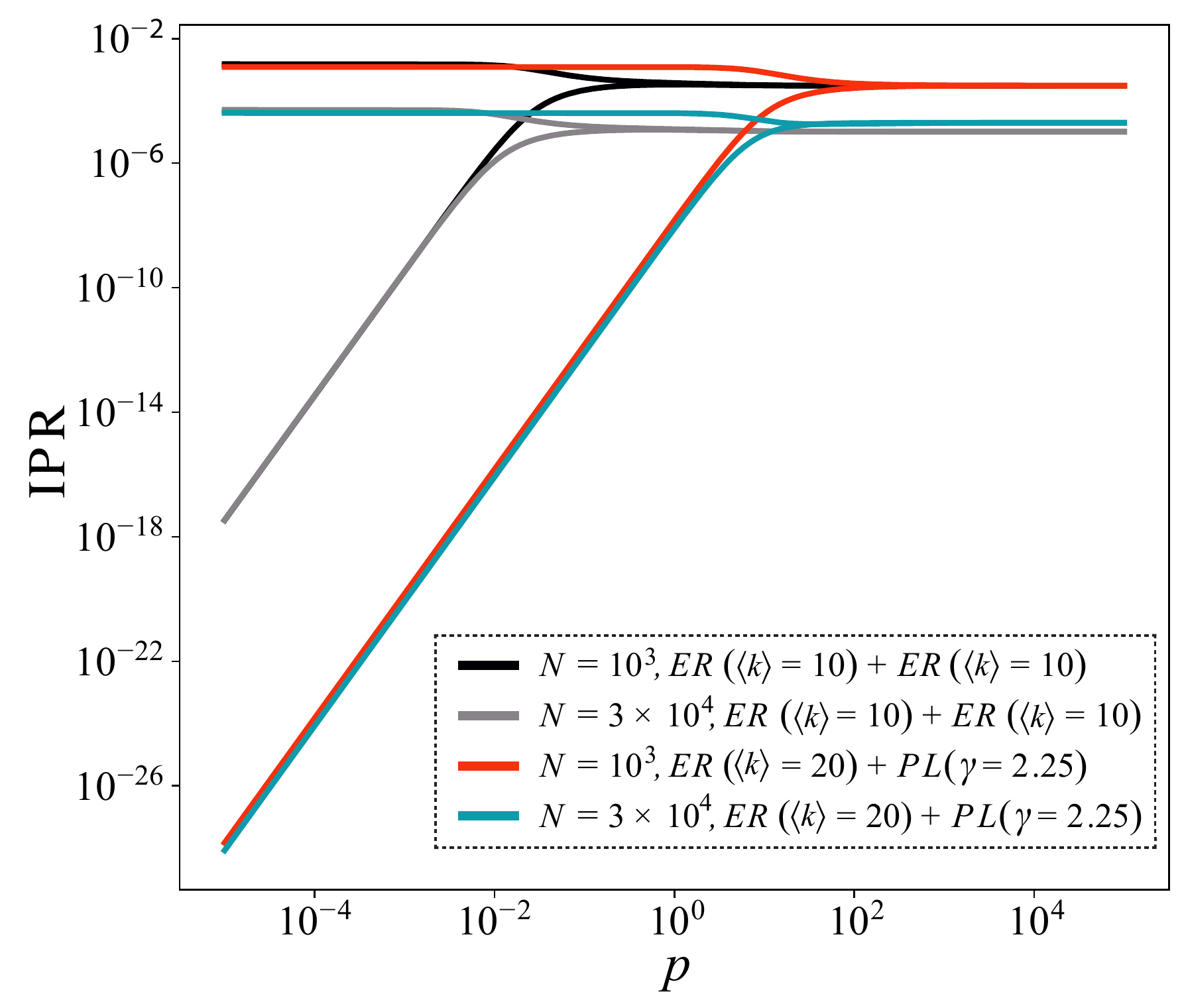}
\includegraphics[width=0.9\columnwidth]{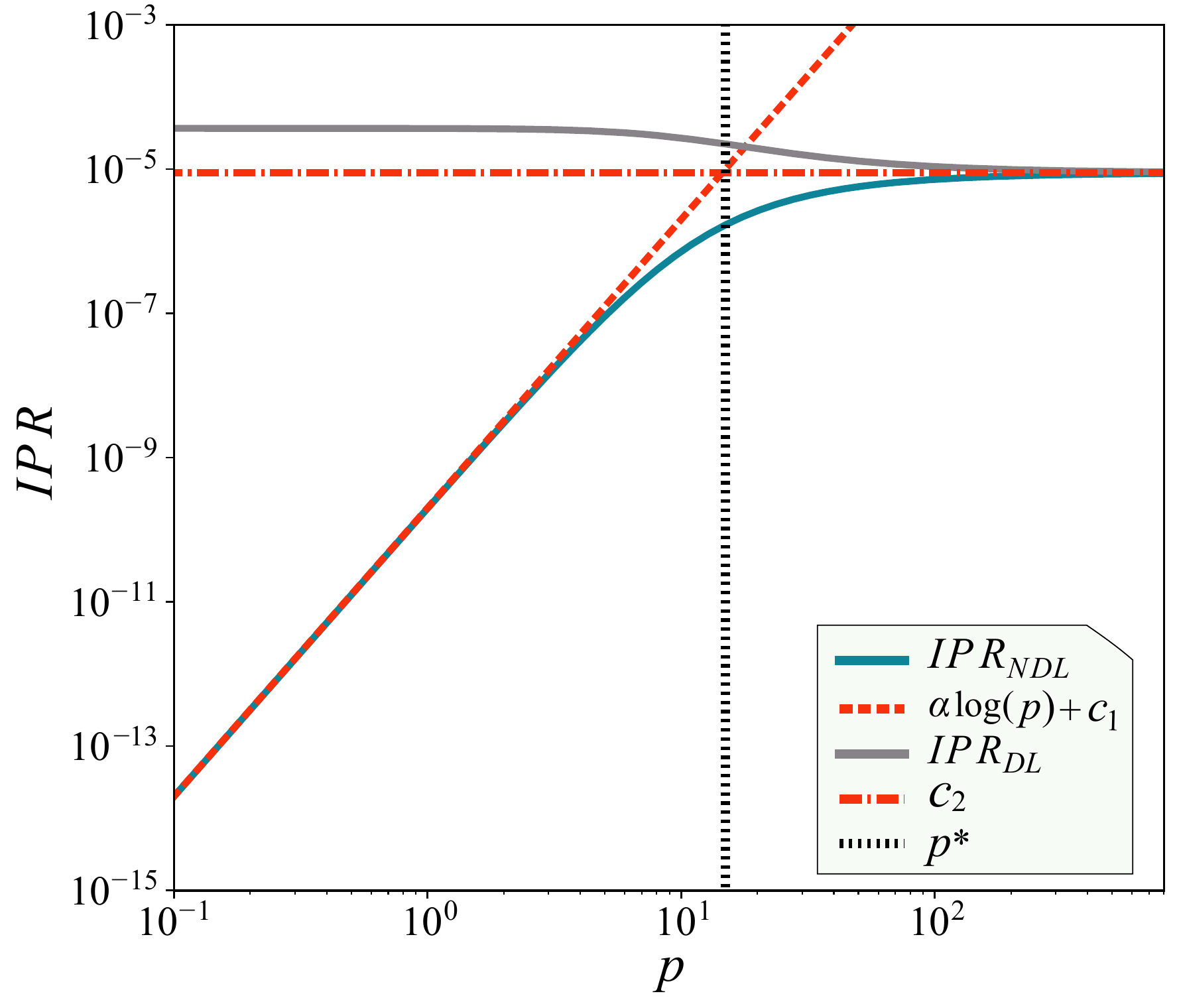}
\caption{The top panel shows the contribution of the dominating and non-dominating network layers to the $\text{IPR}$ as a function of $p$ for several 2-layer network configurations. In the bottom panel, we present an example, where the layer-localized and delocalized regimes are characterized by $\log(\text{IPR}_{NDL}(p<p^*)) \approx \alpha \log(p) + c_1$ and $\log(\text{IPR}_{DL}(p>p^*)) \approx c_2$ (dashed and dot-dashed lines), respectively, and the delocalization transition coupling $p^*$ is given as the crossing of these two curves (vertical dotted line).
Here $\alpha \approx 4$, $c_1 = -9.698$ and $c_2\approx 8.92 \times 10^{-6}$}
\label{Fig:NonCollapsing}
\end{figure}

In network theory~\cite{Boccaletti06:PR,Costa07:AP,Barrat08:book,Newman010:book,Mieghem:2011}, the inverse participation ratio is commonly used to characterize the localization 
features of a network~\cite{Goltsev2013, deArruda2017Feb, Arruda2018}. It is defined as
\begin{equation}
 \text{IPR}(\V) = \sum_i^N \V_i^4.
\end{equation}
Here, $N$ is the network size (here $N=2n$).
Note that, in 2-layer multiplex networks, $\text{IPR}(\V) = \text{IPR}(\V_1) + \text{IPR}(\V_2)$. For 
the sake of notation, we denote the $\text{IPR}$ of the dominating and non-dominating layers as 
$\text{IPR}_{DL}$ and $\text{IPR}_{NDL}$, respectively: $\text{IPR}_{DL}\equiv \text{IPR}(\V_1)$ and $\text{IPR}_{NDL}\equiv \text{IPR}(\V_2)$. 
Furthermore, since $\A$ is a function of 
$p$, both, their eigenvalues and eigenvectors also depend on $p$, therefore, $\text{IPR}\equiv \text{IPR}(p)$.
As a consequence, by tuning $p$ we can observe two different eigenvector regimes characterized by the $\text{IPR}$~\cite{deArruda2017Feb}: (i) layer-localized and (ii) delocalized. 
This statement is exemplified in the top panel of Fig.~\ref{Fig:NonCollapsing} where, without loss of generality, 
we consider homogeneous layers, so we can not observe node localization. When $p\ll p^*$, the states are concentrated in a sub-extensive part of the eigenvector. 
Such part of the eigenvector corresponds to the dominating layer.
However, the density of non-negligible eigenvector components corresponding to the non-dominating layer increases with the coupling parameter $p$.
Indeed, from Fig.~\ref{Fig:NonCollapsing} (top panel), we observe that in the localized regime, $p<p^*$, 
the $\text{IPR}$'s contribution of the non-dominating layer can be characterized by a 
power-law (i.e.~a linear trend in log-log scale), that is, $\log(\text{IPR}_{NDL}(p<p^*)) \approx \alpha \log(p) + c_1$
with $\alpha\approx 4$. 
Moreover, in the delocalized regime, $p>p^*$, 
the states are evenly extended and do not change anymore with $p$ and they are characterized by $\log(\text{IPR}(p>p^*)) \approx c_2$. Therefore, 
the coupling $p^*$, characterizing the delocalization transition, can be defined 
as the value of $p$ such that $\text{IPR}_{NDL}(p<p^*) = \text{IPR}(p>p^*)$. 
This is illustrated in the bottom panel of Fig.~\ref{Fig:NonCollapsing}. With this prescription, we were able to 
systematically characterize the structural transition of the eigenvectors of our 2-layer network by means 
of the $\text{IPR}$. 

\begin{figure}[t!]\centering
\includegraphics[width=0.9\columnwidth]{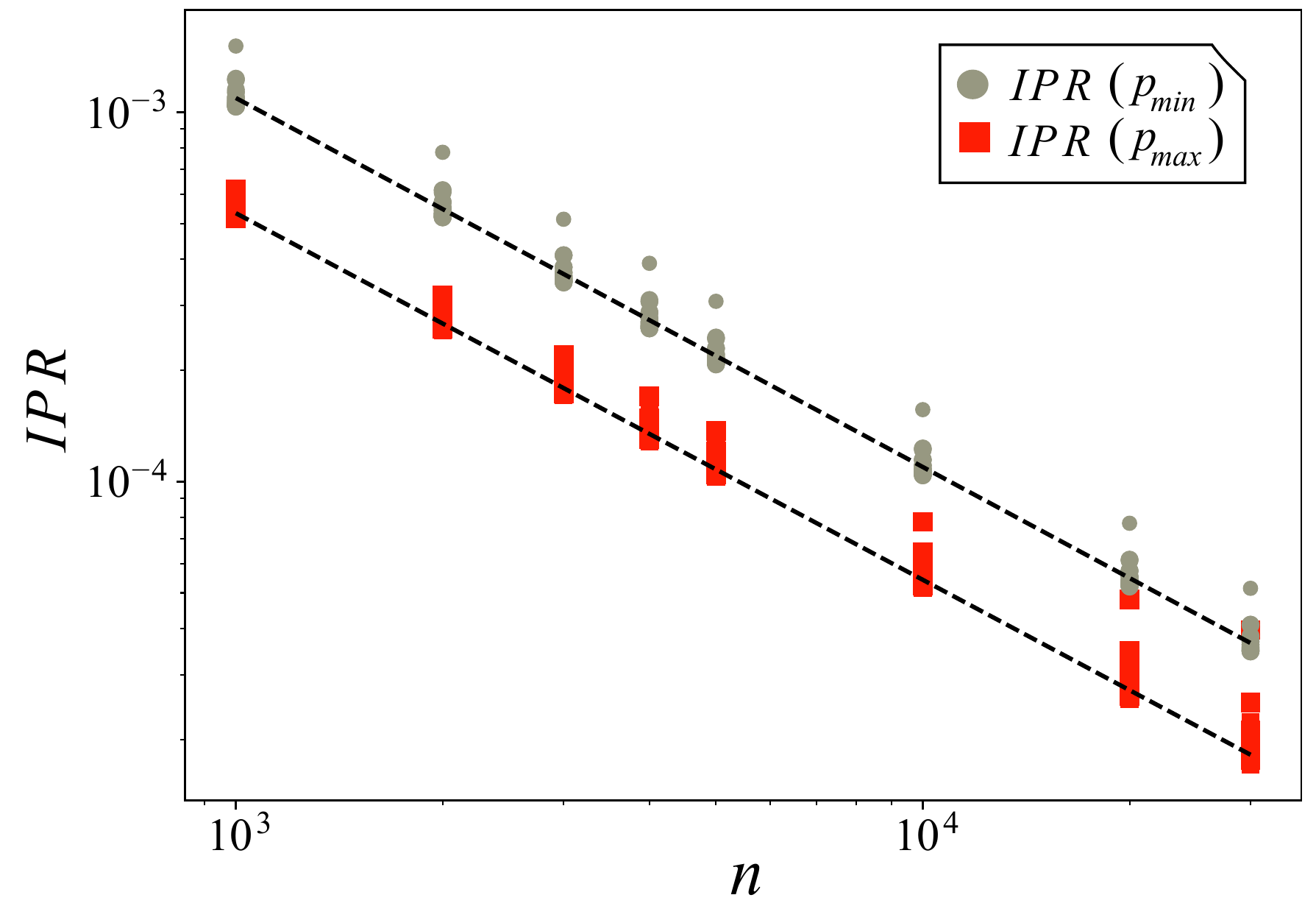}
\includegraphics[width=0.9\columnwidth]{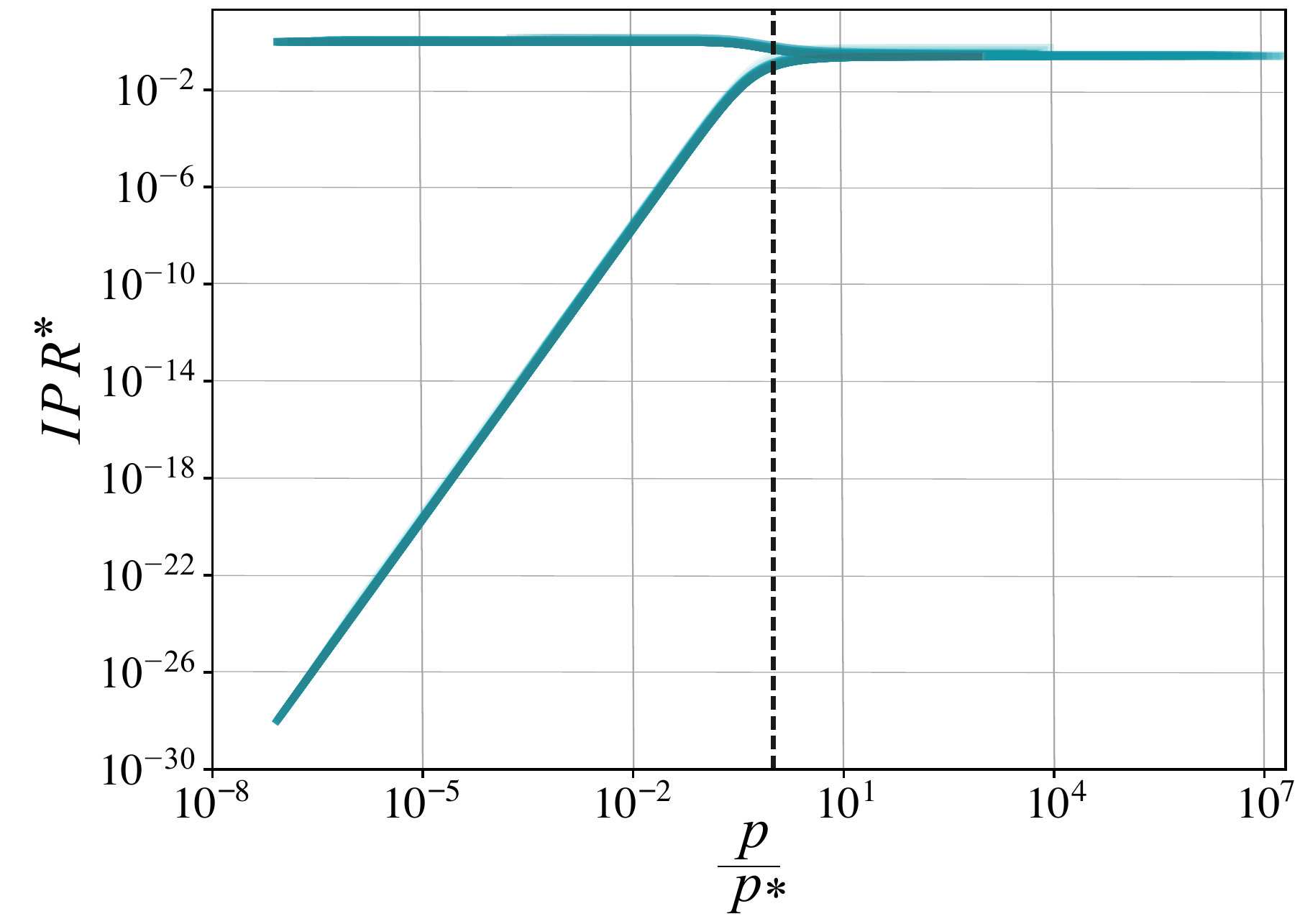}
\caption{The top panel shows the scaling of the $\text{IPR}$ with $n$, $\text{IPR} \sim n^{-\delta}$, for both the minimum and maximum values of $p$ calculated: $\text{IPR}(p_{\min})$ and $\text{IPR}(p_{\max})$. In both cases we estimated $\delta \approx 1$. The network sizes used here are $n = 1 \times 10^3$, $2 \times 10^3$, $3 \times 10^3$, $4 \times 10^3$, $5 \times 10^3$, $10 \times 10^3$, $20 \times 10^3$, and $30 \times 10^3$. The bottom panel represents the collapsing of the curves $\text{IPR}$ vs.~$p$ by the use of $\text{IPR}^* = n \times \text{IPR}$ (see top figure) and $p' = p/p^*$. In the homogeneous layer configurations (ER networks) we used all the combinations of $k_1 = \{ 10, 20, 30, 40, 50, 60, 70, 80, 90, 100\}$ and $k_2 = \{ 10, 20, 30, 40\}$. In the mixed case, where one layer is an ER network and the other is a PL network, we considered all the combinations of $\gamma_1 = \{2.25, 2.5, 3.5\}$ for the PL layer and $k_2 = \{ 20, 30, 40\}$ for the ER layer.
}
\label{Fig:Collapsing}
\end{figure}

Importantly, Fig.~\ref{Fig:NonCollapsing} also shows that different network configurations produce 
different $\text{IPR}$ functions. Note that, although all the curves have a similar behavior, they are 
shifted in both axis. Aside from the dependence of the $\text{IPR}$ on $p$, it also depends on the system size as $\text{IPR}(p,n) \sim n^{-\delta}$; as shown in the top panel of Fig.~\ref{Fig:Collapsing}.
In fact, as discussed in~\cite{deArruda2017Feb, Arruda2018}, the 
layer-localized regime scales as $\text{IPR} (p<p^*) \sim O\left(\frac{1}{{m}} \right)$, while in the delocalized 
regime it behaves as $\text{IPR} (p>p^*) \sim O\left(\frac{1}{{nm}} \right)$. Thus, for a fixed number of 
layers and in the thermodynamic limit, both regimes scale similarly with $\delta \approx 1$ (see Fig.~\ref{Fig:Collapsing}, top panel).

As a consequence of the $\text{IPR}$ behavior described above, and taking into account both the dependencies on the system size $n$ and on the coupling parameter $p$ (see also~\cite{deArruda2017Feb, Arruda2018}), we define
\begin{eqnarray}
 \text{IPR}^* &=& n \times \text{IPR}, \label{eq:ipr_star}  \\
 p^* &=& \frac{p}{p'}. \label{eq:p_prime}
\end{eqnarray}
Under these {\it scalings}, all the $\text{IPR}^*$ vs.~$p'$ curves should collapse on top of a {\it universal}
curve. Indeed, the scaling of the $\text{IPR}$ is shown in the bottom panel of Fig.~\ref{Fig:Collapsing}. 

\begin{figure}[t!]\centering
\includegraphics[width=0.95\columnwidth]{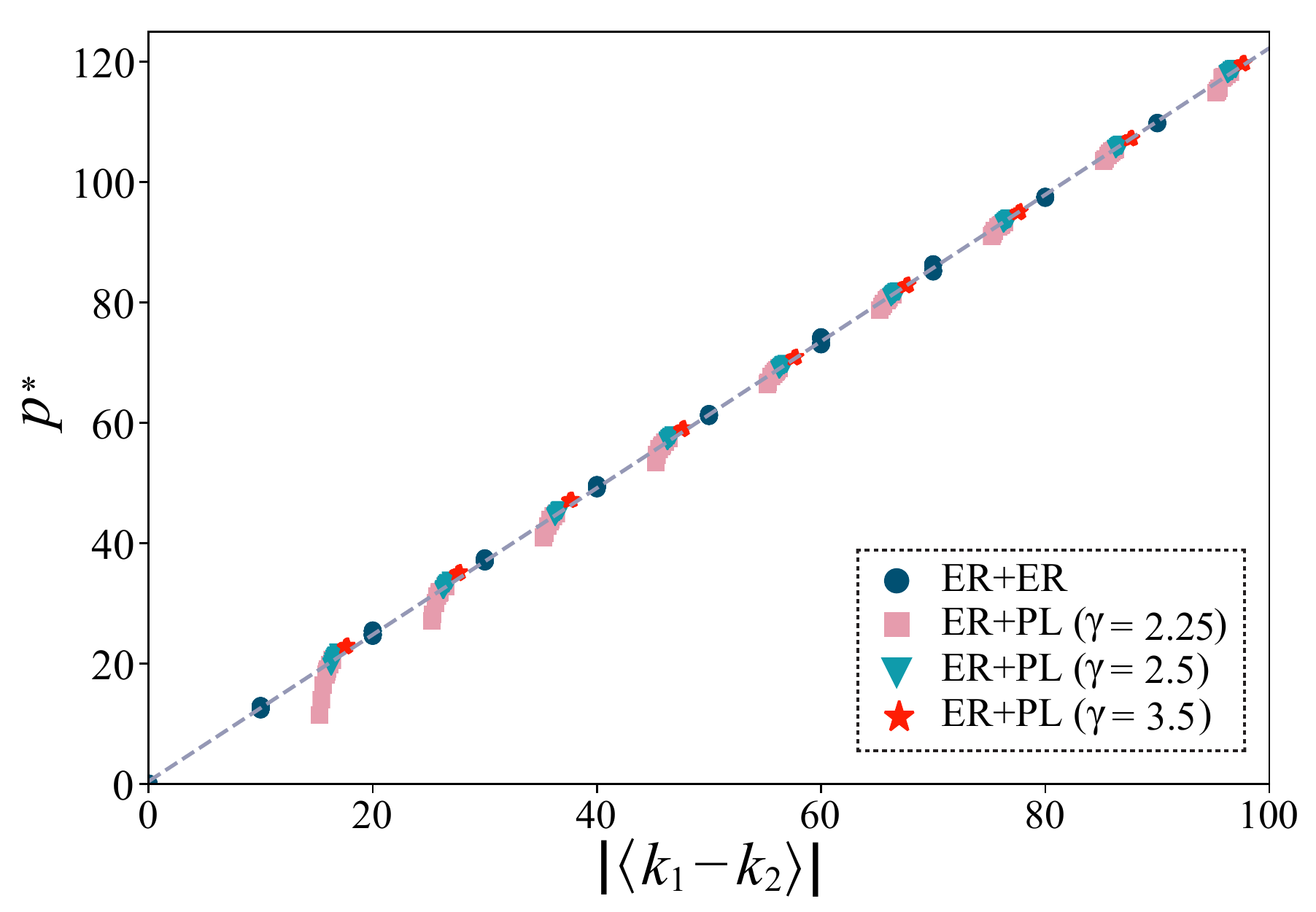}
\caption{Delocalization transition coupling $p^*$ as a function of the absolute degree 
difference of the two layers forming the multiplex. The dashed line is
$p^* = \beta_1 \langle k_1 - k_2 \rangle + \beta_2$, with $\beta_1 = 1.218$ and 
$\beta_2 = 0.430$. Same network parameters as in Fig.~\ref{Fig:Collapsing}.} 
\label{Fig:Pstar}
\end{figure}

It is relevant to stress that the quantity driving the $\text{IPR}$ scaling, for a fixed $p$ and fixed layer 
structure (i.e.,~the class of network considered), is the network size. For example, the $\text{IPR}$ of 
an Erd\"os-R\'enyi (ER) single-layer network scales as $n^{-1}$ and node-localization is 
absent~\cite{Satorras2016}. Furthermore, for power-law (PL) networks, $P(k) \sim k^{-\gamma}$, 
depending on the value of $\gamma$ one can observe different scaling laws that depend on $n$; 
namely, the network can present $k$-core hub localization~\cite{Satorras2016}. In multiplex 
networks, the layer-localization phenomena was already discussed in~\cite{deArruda2017Feb}, in the 
context of disease localization. There, it was shown that the states can be localized in 
one or more layers. Moreover, in the multiplex case, the scaling is not a universal property with respect to different layer configurations. 
For instance, for a fixed value of $p$, by changing the average degree of a layer in the multiplex, the 
eigenstates can transit from localization to delocalization (the opposite is also true); see for example 
Fig.~\ref{Fig:NonCollapsing}. 
Thus, the universality shown in Fig.~\ref{Fig:Collapsing} could not be robust against the inner configuration of the layers. A detailed study of the $\text{IPR}$ as a function of $p$, allows relating the structural properties of the multiplex with the delocalization transition coupling $p^*$.
Interestingly enough, we found that $p^*$ is approximately linearly described by the difference between the average degree of the layers; that is
\begin{equation} \label{eq:p_star}
 p^* = \beta_1 |\langle k_1 - k_2 \rangle| + \beta_2 ,
\end{equation}
as can be clearly seen in Fig.~\ref{Fig:Pstar} for a wide range of parameter combinations.
This relationship describes the change of the $\text{IPR}$ curves with the inner configuration of the layers. Surprisingly, despite the local structures that might appear inside the multiplex (cycles), for the range of parameters studied, the average degree difference describes reasonably well the eigenvector structural transition.

To the best of our knowledge, a definition of the eigenvector structural transition, from layer-localization to delocalization, is not available in the literature. This concept might seem natural at first glance, however, since there is no abrupt transition on the spectral properties of the adjacency matrix nor a singularity to look for, a transition in the strict sense is hard to define. In the following, we formalize this transition and provide a definition for it. It is important to stress that this structural change, characterized by the $\text{IPR}$, is continuous and the $\text{IPR}$ does not vanish in the thermodynamic limit. We also remark, however, that an abrupt structural transition can be found in the Laplacian matrix \cite{Radicchi2014,Cozzo2016}. Moreover, the adjacency and the Laplacian matrices concern different dynamical processes. So, our interest here is the adjacency matrix due to its relevance for the SIS epidemic spreading process.

In~\cite{deArruda2017Feb} it was shown that the disease spreading, more precisely the SIS model, on a multilayer network might present a transition from layer-localization to delocalization. This phenomenon depends on the layer configurations, as well as the spreading parameters~\cite{deArruda2017Feb}. Motivated by~\cite{Goltsev2013}, the authors extended the concept of localization from node localization to layer-localization. Here, by establishing $\text{IPR}^*$ as in Eq.~(\ref{eq:ipr_star}), we go one step farther since this quantity does not vanish in the thermodynamic limit. Notably, the most important contribution of our analysis regards the nature of the SIS transition on multiplex networks. From the quenched mean-field (QMF) theory, where we assume that the individual probabilities are independent, (see~\cite{deArruda2017Feb, Arruda2018}) we have
\begin{equation}
\frac{d}{dt}
\left[
\begin{array}{c}
 \Y_1 \\
 \hline
 \Y_2
\end{array}
\right] =
-\delta
\left[
\begin{array}{c}
 \Y_1 \\
 \hline
 \Y_2
\end{array}
\right]
+ \lambda
 \left[
\begin{array}{c|c}
\A_1 & \frac{\eta}{\lambda}\I \\
\hline
\frac{\eta}{\lambda}\I & \A_2
\end{array}
\right]
\left[
\begin{array}{c}
 \Y_1 \\
 \hline
 \Y_2
\end{array}
\right] 
+ O(y^2),
\end{equation}
where the transition from a disease-free state to an endemic state occurs at the critical point given by $\lambda_c = \delta/\Lambda_1$, with $\Lambda_1$ being the leading eigenvalue of $\A$, see Eq.~(\ref{eq:definition}), and $p = \eta/\lambda$ in our context. Therefore, $p^* = \eta/\lambda_c = \eta\Lambda_1$. Moreover, we know that if $p > p^*$ the eigenvectors of the multiplex are in the delocalized regime. Conversely, if $p < p^*$ the eigenvectors show layer localization. We can now translate this condition into the SIS epidemic spreading context. 
Thus, from the QMF theory, if
\begin{equation} \label{eq:eta}
 \eta > \frac{p^*}{\Lambda_1},
\end{equation}
where $\eta$ is the inter-layer spreading rate, the disease is delocalized, and the whole multiplex is active. Notice that if Eq.~(\ref{eq:eta}) is not satisfied a transition from a disease-free state to a layer-localized state is still present. Aside from that, note that the evaluation of Eq.~(\ref{eq:eta}) is not trivial since $p^*$ and $\Lambda_1$ depend on $\eta$. We also remark that, since the delocalization {\it transition} is continuous, i.e.,~it is not characterized by a divergence on a given derivative (see for instance Fig.~\ref{Fig:Collapsing}, where the finite size effects are not present and the curves for different network sizes collapse), the corresponding dynamics is also expected to suffer a smooth transition. This is very relevant, because it might be easiest to study localization-delocalization transitions through spectra properties than the nature of a dynamical transition, which usually involves more refined and cumbersome numerical techniques. We also want to stress that an important consequence of the renormalization defined in Eqs.~(\ref{eq:ipr_star}) and~(\ref{eq:p_prime}) is that they allow for the analysis of finite systems. In other words, these definitions do not require the thermodynamic limit. For the sake of rigor, however, we remark that a true critical point is properly defined only in the infinite size limit.

\begin{figure}[t!]\centering
\includegraphics[width=\columnwidth]{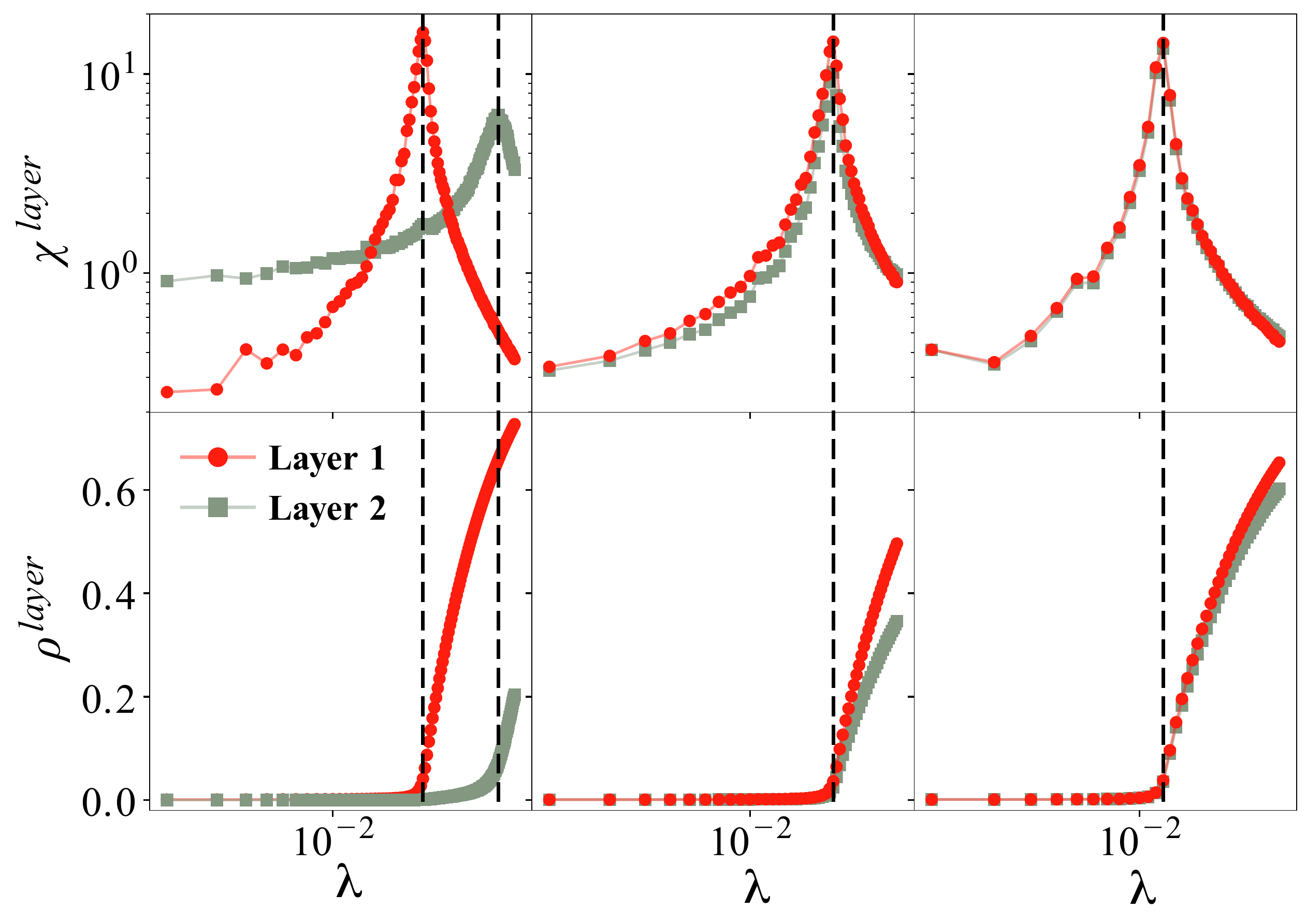}
\caption{
Quasi stationary simulation of a SIS epidemic spreading. 
Susceptibility $\chi$ (upper panels) and order parameter $\rho$ (lower panels) as a function of $\lambda$, for each layer. 
The multiplex network used in this simulation is composed by two ER networks with 
$\langle k \rangle = 30$ and $\langle k \rangle = 10$. 
From the left to the right we have three different structural regimes: 
(i) the layer-localized regime with $\eta = 0.01$, 
(ii) near the structural delocalization transition point, 
$\eta \Lambda_1 \approx p^*$ with $\eta = 0.85$, $p^* = 24.795$ [from Eq.~(\ref{eq:p_star})] and 
$\Lambda_1 = 31.058$, and (iii) the delocalized regime with $\eta = 3.0$.
The gray dashed lines mark the susceptibility peaks.}
\label{Fig:QS}
\end{figure}

Furthermore, in Fig.~\ref{Fig:QS} we present Monte Carlo simulations using the quasi-stationary algorithm~\cite{Oliveira2005, Silvio2012, deArruda2017Feb, Arruda2018}, where the absorbing state is avoided, conferring a numerical validation to our analysis. We evaluate the dynamical response of the SIS model on the different structural regimes: (i) the layer-localized regime, $\eta \Lambda_1 < p^*$, (ii) near the structural transition point, $\eta \Lambda_1 \approx p^*$ and (iii) the delocalized regime, $\eta \Lambda_1 > p^*$. 
Note that, in the layer-localized regime, the epidemics plays a role similar to an external field for the non-dominating layer. Thus, the non-dominating layer plays a minor role. This behavior can be observed in both, the order parameter (for each layer) $\rho$, and the susceptibility (also calculated individually for each layer) $\chi$, see the left panels of Fig.~\ref{Fig:QS}.
In the susceptibility curves we can even see a second peak, as predicted in~\cite{deArruda2017Feb}. 
As discussed before, the eigenvector transition from layer-localization to delocalization is not a sharp transition. This is also illustrated in the middle panels of Fig.~\ref{Fig:QS}. In the transition regime, a small change in $\eta$ does not imply an important change of behavior. Note that at the critical point (peak of susceptibility) the fraction of infected individuals, $\rho$, is similar in both layers. Besides, these two curves have different growth rates for larger values of $\lambda$. Finally, in the right panels of Fig.~\ref{Fig:QS}, we present the delocalized structural regime, where the curves of $\chi$ and $\rho$ are practically the same for both layers. In this regime, as we increase $\eta$, we are also increasing the leading eigenvalue of $\A$ and, thus, moving the critical point to the left. Therefore for large $\eta$, using perturbative analysis, we can interpret the block diagonal matrices as a perturbation on the off-diagonal ones.

% conclusions

In this paper we have formalized the layer-localization to delocalization transition in bilayer multiplex networks. This process was already anticipated in~\cite{deArruda2017Feb}, however a proper definition was lacking. In the latter study, the different structural regimes (layer-localized and delocalized) were characterized, but the transition point between them was not defined. In order to properly define the transition, we first performed a scaling analysis of the $\text{IPR}$ of the eigenvectors of the adjacency matrix of the bilayer multiplex network as a function of the parameter associated to the coupling between the layers. Furthermore, we also found a linear relationship between the delocalization transition point, $p^*$, and the difference of the average degree between the layers. We remark that this relationship is valid for the set of parameters evaluated here. That is, for homogeneous-layer settings and mixed-layer settings both with a reasonably high average degree, since in order to observe layer-localization, layer dominance is needed. 

Finally, we have also applied our results on the universality of layer delocalization to disease spreading. By using the QMF approach, we were able to define a criterion for disease layer-localization, which was validated through Monte Carlo simulations. This constitutes a step towards a better understanding of the delocalization transition reported for disease dynamics on multilayer networks~\cite{deArruda2017Feb}. We hope that our work could also motivate further research on the impact of the layer delocalization transition in dynamical processes, as well as on the universality of other properties of multiplex networks.

\begin{acknowledgments}
G. F. A. and Y. M. acknowledge partial support from Intesa Sanpaolo Innovation Center. J.A.M.-B. thanks support from
FAPESP (Grant No.~2019/06931-2), Brazil, and PRODEP-SEP (Grant No.~511-6/2019.-11821), Mexico. F.A.R acknowledge support from CNPq (Grant No. 309266/2019-0) and Fapesp (Grant No. 13/07375-0). Y. M. acknowledges partial support from the Government of Arag\'on, Spain through grant E36-17R, and by MINECO and FEDER funds (Grant No.~FIS2017-87519-P). Research carried out using the computational resources of the Center for Mathematical Sciences Applied to Industry (CeMEAI) funded by FAPESP (Grant No.~2013/07375-0). The funders had no role in study design, data collection, and analysis, decision to publish, or preparation of the manuscript. 
\end{acknowledgments}

%\bibliographystyle{apsrev}
%\bibliography{references}

\begin{thebibliography}{99}

\bibitem{Stanley1971} H. E. Stanley. Introduction to Phase Transitions and Critical Phenomena. Oxford University Press, New York, 1971.

\bibitem{Marro1999} J. Marro, and R. Dickman. Nonequilibrium Phase Transitions in Lattice Models. Cambridge University Press, Cambridge, UK, 1999. doi:10.1017/CBO9780511524288

\expandafter\ifx\csname natexlab\endcsname\relax\def\natexlab#1{#1}\fi
\expandafter\ifx\csname bibnamefont\endcsname\relax
  \def\bibnamefont#1{#1}\fi
\expandafter\ifx\csname bibfnamefont\endcsname\relax
  \def\bibfnamefont#1{#1}\fi
\expandafter\ifx\csname citenamefont\endcsname\relax
  \def\citenamefont#1{#1}\fi
\expandafter\ifx\csname url\endcsname\relax
  \def\url#1{\texttt{#1}}\fi
\expandafter\ifx\csname urlprefix\endcsname\relax\def\urlprefix{URL }\fi
\providecommand{\bibinfo}[2]{#2}
\providecommand{\eprint}[2][]{\url{#2}}

\bibitem[{\citenamefont{M\'endez-Berm\'udez
  et~al.}(2017)\citenamefont{M\'endez-Berm\'udez, de~Arruda, Rodrigues, and
  Moreno}}]{Bermudez2017}
\bibinfo{author}{\bibfnamefont{J.~A.} \bibnamefont{M\'endez-Berm\'udez}},
  \bibinfo{author}{\bibfnamefont{G.~F.} \bibnamefont{de~Arruda}},
  \bibinfo{author}{\bibfnamefont{F.~A.} \bibnamefont{Rodrigues}},
  \bibnamefont{and} \bibinfo{author}{\bibfnamefont{Y.}~\bibnamefont{Moreno}},
  \bibinfo{journal}{Phys. Rev. E} \textbf{\bibinfo{volume}{96}},
  \bibinfo{pages}{012307} (\bibinfo{year}{2017}).

\bibitem[{\citenamefont{S{\'a}nchez-Garc\'{\i}a
  et~al.}(2014)\citenamefont{S{\'a}nchez-Garc\'{\i}a, Cozzo, and
  Moreno}}]{Cozzo2014}
\bibinfo{author}{\bibfnamefont{R.~J.} \bibnamefont{S{\'a}nchez-Garc\'{\i}a}},
  \bibinfo{author}{\bibfnamefont{E.}~\bibnamefont{Cozzo}}, \bibnamefont{and}
  \bibinfo{author}{\bibfnamefont{Y.}~\bibnamefont{Moreno}},
  \bibinfo{journal}{Phys. Rev. E} \textbf{\bibinfo{volume}{89}},
  \bibinfo{pages}{052815} (\bibinfo{year}{2014}).

%\bibitem[{\citenamefont{Cozzo et~al.}(2016)\citenamefont{Cozzo, de~Arruda,
%  Rodrigues, and Moreno}}]{cozzo2016multilayer}
%\bibinfo{author}{\bibfnamefont{E.}~\bibnamefont{Cozzo}},
%  \bibinfo{author}{\bibfnamefont{G.~F.} \bibnamefont{de~Arruda}},
%  \bibinfo{author}{\bibfnamefont{F.~A.} \bibnamefont{Rodrigues}},
%  \bibnamefont{and} \bibinfo{author}{\bibfnamefont{Y.}~\bibnamefont{Moreno}},
%  in \emph{\bibinfo{booktitle}{Interconnected Networks}}
%  (\bibinfo{publisher}{Springer}, \bibinfo{year}{2016}), pp.
%  \bibinfo{pages}{17--35}.

\bibitem[{\citenamefont{de~Arruda et~al.}(2017)\citenamefont{de~Arruda, Cozzo,
  Peixoto, Rodrigues, and Moreno}}]{deArruda2017Feb}
\bibinfo{author}{\bibfnamefont{G.~F.} \bibnamefont{de~Arruda}},
  \bibinfo{author}{\bibfnamefont{E.}~\bibnamefont{Cozzo}},
  \bibinfo{author}{\bibfnamefont{T.~P.} \bibnamefont{Peixoto}},
  \bibinfo{author}{\bibfnamefont{F.~A.} \bibnamefont{Rodrigues}},
  \bibnamefont{and} \bibinfo{author}{\bibfnamefont{Y.}~\bibnamefont{Moreno}},
  \bibinfo{journal}{Phys. Rev. X} \textbf{\bibinfo{volume}{7}},
  \bibinfo{pages}{011014} (\bibinfo{year}{2017}), ISSN
  \bibinfo{issn}{2160-3308}.

\bibitem[{\citenamefont{Cozzo et~al.}(2018)\citenamefont{Cozzo, {De Arruda},
  Rodrigues, and Moreno}}]{cozzo2018multiplex}
\bibinfo{author}{\bibfnamefont{E.}~\bibnamefont{Cozzo}},
  \bibinfo{author}{\bibfnamefont{G.~F.} \bibnamefont{{De Arruda}}},
  \bibinfo{author}{\bibfnamefont{F.~A.} \bibnamefont{Rodrigues}},
  \bibnamefont{and} \bibinfo{author}{\bibfnamefont{Y.}~\bibnamefont{Moreno}},
  \emph{\bibinfo{title}{Multiplex networks: basic formalism and structural
  properties}} (\bibinfo{publisher}{Springer}, \bibinfo{year}{2018}).

\bibitem[{\citenamefont{de~Arruda
  et~al.}(2018{\natexlab{a}})\citenamefont{de~Arruda, Cozzo, Rodrigues, and
  Moreno}}]{deArruda2018}
\bibinfo{author}{\bibfnamefont{G.~F.} \bibnamefont{de~Arruda}},
  \bibinfo{author}{\bibfnamefont{E.}~\bibnamefont{Cozzo}},
  \bibinfo{author}{\bibfnamefont{F.~A.} \bibnamefont{Rodrigues}},
  \bibnamefont{and} \bibinfo{author}{\bibfnamefont{Y.}~\bibnamefont{Moreno}},
  \bibinfo{journal}{New Journal of Physics} \textbf{\bibinfo{volume}{20}},
  \bibinfo{pages}{095004} (\bibinfo{year}{2018}{\natexlab{a}}).
  
\bibitem{Kivela2014}
M. Kivel\"a, A. Arenas, M. Barthelemy, J. P. Gleeson, Y. Moreno, and M. A. Porter,
%Multilayer networks,
J. Complex Networks {\bf 2}, 203 (2014).

\bibitem{BoccalettiPR2014}
S. Boccaletti, G. Bianconi, R. Criado, C. I. del Genio, J. G\'omez-Garde\~nes, M. Romance, 
I. Sendi\~na-Nadal, Z. Wang, M. Zanin, 
%%The structure and dynamics of multilayer networks, 
Phys. Rep. {\bf 544}, 1 (2014).

\bibitem{Bianconi2018} G. Bianconi. Multilayer Networks: Structure and Function. Oxford University Press, Oxford, 2018.

\bibitem{Aleta2019} A. Aleta and Y. Moreno, Annual Reviews of Condensed Matter Physics {\bf 10}, 45 (2019).

\bibitem[{\citenamefont{Cozzo et~al.}(2013)\citenamefont{Cozzo,
  Ba{\ifmmode\tilde{n}\else{\~n}\fi}os, Meloni, and Moreno}}]{Cozzo2013Nov}
\bibinfo{author}{\bibfnamefont{E.}~\bibnamefont{Cozzo}},
  \bibinfo{author}{\bibfnamefont{R.~A.}
  \bibnamefont{Ba{\ifmmode\tilde{n}\else{\~n}\fi}os}},
  \bibinfo{author}{\bibfnamefont{S.}~\bibnamefont{Meloni}}, \bibnamefont{and}
  \bibinfo{author}{\bibfnamefont{Y.}~\bibnamefont{Moreno}},
  \bibinfo{journal}{Phys. Rev. E} \textbf{\bibinfo{volume}{88}},
  \bibinfo{pages}{050801} (\bibinfo{year}{2013}).
  
  \bibitem{Boccaletti06:PR}
S.~Boccaletti, V.~Latora, Y.~Moreno, M.~Chavez, D.~Hwang, Physics Reports 424~(4) (2006) 175--308.

\bibitem{Costa07:AP}
L.~Costa, F.~Rodrigues, G.~Travieso, P.~Boas, Advances in Physics 56~(1) (2007)
  167--242.
  
 \bibitem{Barrat08:book}
A.~Barrat, M.~Barthlemy, A.~Vespignani. Dynamical processes on complex
  networks, Cambridge University Press New York, NY, USA, 2008.

\bibitem{Newman010:book}
M.~Newman. Networks: an introduction. Oxford University Press, Inc., 2010.

\bibitem{Mieghem:2011} P.~V. Mieghem. Graph Spectra for Complex Networks. Cambridge University
  Press, New York, NY, USA, 2011.

\bibitem[{\citenamefont{Goltsev et~al.}(2012)\citenamefont{Goltsev,
  Dorogovtsev, Oliveira, and Mendes}}]{Goltsev2013}
\bibinfo{author}{\bibfnamefont{A.~V.} \bibnamefont{Goltsev}},
  \bibinfo{author}{\bibfnamefont{S.~N.} \bibnamefont{Dorogovtsev}},
  \bibinfo{author}{\bibfnamefont{J.~G.} \bibnamefont{Oliveira}},
  \bibnamefont{and} \bibinfo{author}{\bibfnamefont{J.~F.~F.}
  \bibnamefont{Mendes}}, \bibinfo{journal}{Phys. Rev. Lett.}
  \textbf{\bibinfo{volume}{109}}, \bibinfo{pages}{128702}
  (\bibinfo{year}{2012}).

\bibitem[{\citenamefont{de~Arruda
  et~al.}(2018{\natexlab{b}})\citenamefont{de~Arruda, Rodrigues, and
  Moreno}}]{Arruda2018}
\bibinfo{author}{\bibfnamefont{G.~F.} \bibnamefont{de~Arruda}},
  \bibinfo{author}{\bibfnamefont{F.~A.} \bibnamefont{Rodrigues}},
  \bibnamefont{and} \bibinfo{author}{\bibfnamefont{Y.}~\bibnamefont{Moreno}},
  \bibinfo{journal}{Physics Reports} \textbf{\bibinfo{volume}{756}},
  \bibinfo{pages}{1} (\bibinfo{year}{2018}).
 % {\natexlab{b}}), ISSN \bibinfo{issn}{0370-1573}
  % \bibinfo{note}{Fundamentals of spreading processes in single and multilayer complex networks}.

\bibitem[{\citenamefont{Pastor-Satorras and Castellano}(2016)}]{Satorras2016}
\bibinfo{author}{\bibfnamefont{R.}~\bibnamefont{Pastor-Satorras}}
  \bibnamefont{and}
  \bibinfo{author}{\bibfnamefont{C.}~\bibnamefont{Castellano}},
  \bibinfo{journal}{Scientific Reports} \textbf{\bibinfo{volume}{6}},
  \bibinfo{pages}{18847} (\bibinfo{year}{2016}).
  
  \bibitem{Radicchi2014} F. Radicchi, Phys. Rev. X {\bf 4}, 021014 (2014)

\bibitem{Cozzo2016} E. Cozzo, and Y. Moreno,
%Characterization of multiple topological scales in multiplex networks through supra-Laplacian eigengaps, 
Physical Review E {\bf 94}, 052318 (2016)

\bibitem[{\citenamefont{de~Oliveira and Dickman}(2005)}]{Oliveira2005}
\bibinfo{author}{\bibfnamefont{M.~M.} \bibnamefont{de~Oliveira}}
  \bibnamefont{and} \bibinfo{author}{\bibfnamefont{R.}~\bibnamefont{Dickman}},
  \bibinfo{journal}{Phys. Rev. E} \textbf{\bibinfo{volume}{71}},
  \bibinfo{pages}{016129} (\bibinfo{year}{2005}).

\bibitem[{\citenamefont{Ferreira et~al.}(2012)\citenamefont{Ferreira,
  Castellano, and Pastor-Satorras}}]{Silvio2012}
\bibinfo{author}{\bibfnamefont{S.~C.} \bibnamefont{Ferreira}},
  \bibinfo{author}{\bibfnamefont{C.}~\bibnamefont{Castellano}},
  \bibnamefont{and}
  \bibinfo{author}{\bibfnamefont{R.}~\bibnamefont{Pastor-Satorras}},
  \bibinfo{journal}{Phys. Rev. E} \textbf{\bibinfo{volume}{86}},
  \bibinfo{pages}{041125} (\bibinfo{year}{2012}).

\end{thebibliography}

\end{document}